\documentclass[12pt,preprint]{aastex}
\shorttitle{A Brightness Enhancement in Q0957+561}
\shortauthors{Oscoz et al.}
\begin{document}

\title{A Large Brightness Enhancement of the QSO 0957+561 A
Component}

\author{A. Oscoz\altaffilmark{1}, D. Alcalde\altaffilmark{1}, 
M. Serra--Ricart\altaffilmark{1}, E. Mediavilla\altaffilmark{1}, 
J. A. Mu\~noz\altaffilmark{1}}
\affil{Instituto de Astrof\'\i sica de Canarias, E-38205 La Laguna,
Tenerife, Spain}
\email{aoscoz@ll.iac.es, dalcalde@ll.iac.es, mserra@ot.iac.es, 
emg@ll.iac.es, jmunoz@ll.iac.es}

\begin{abstract}
We report an increase of more than 0.2 mag in the optical 
brightness of the leading image (A) of the gravitational lens 
Q0957+561, detected during the 09/2000 -- 06/2001 monitoring 
campaign (2001 observing season). The brightening is similar to 
or even greater than the largest change ever detected during 
the 20 years of monitoring of this system. We discuss two 
different provisional explanations to this event: intrinsic 
source variability or microlensing (either short timescale 
microlensing or cessation of the historical microlensing). An 
exhaustive photometric monitoring of Q0957+561 is needed until 
summer of 2002 and during 2003 to discriminate between these 
possibilities.
\end{abstract}


\section{Introduction}

The first gravitational lens system discovered, Q0957+561 
\citep{wal79}, has become the most observed gravitational 
mirage. This system has been the target of continuous 
monitoring in optical and radio wavelenghts. The early works by 
\citet{flo84,sch86,leh89,van89,sch90}; and \citet{rob91}, were 
followed by other recent monitoring, as those by 
\citet{bes92,kun95,osc96,kun97,osc97,haa99}; and \citet{ser99}.

Three outstanding events can be noticed during these twenty 
years of monitoring. Firstly, the existence of a large 
timescale microlensing (of several years). The analysis made by 
\citet{pel98} with data corresponding to the period 1979--1996 
clearly shows its presence. Next, a strong 0.13--mag intrinsic 
brightening of Q0957+561 in two months was detected by 
\citet{kun95}. And finally, \citet{sch96} noticed a possible 
microlensing event with maximum amplitude of 0.05 mag and a 
timescale of 90 days. The sharp drop detected by Kundic 
et al. allowed to solve the long-standing problem concerning 
the ``short" ($\sim 410$ days) and ``long" ($\sim530$ days) 
time delays between the A and B components of the system. The 
observations confirmed that the short value was the correct one 
\citep{osc96,kun97}, constraining the time delay between 410 
and 440 days. Moreover, this feature allowed to obtain a first 
accurate value for the delay (417$\pm$3 days, Kundic et al. 1997; 
424$\pm$3 days, Oscoz et al. 1997). This robust estimate lead 
to search for the existence of possible microlensing events 
\citep{gil98,goi98}, but no other event of the type reported by 
\citet{sch96} has been detected (see Gil--Merino et al. 2001).

The Instituto de Astrof\'\i sica de Canarias gravitational 
lensing group started a long-term monitoring program on this 
system in 1996, with the 0.82-m IAC80 telescope at the 
Observatorio del Teide in Tenerife, Spain. Our set of almost 
500 individual observations in $R$ band, together with several 
hundreds of points in $V$ band, constitutes one of the largest 
photometric database of a gravitational lens system. The 
application of a new data reduction method (to improve the 
original aperture photometry), and the development of a new 
procedure to estimate the time delay, gave a  
value of $\Delta t_{Q0957} = (425\pm4)$ days \citep{ser99}. The 
accuracy in the time delay was improved furthermore by 
including data from other groups in the period 1984--1999 and 
by using several statistical methods for the calculations. A 
new value of $422.6\pm0.6$ days was derived \citep{osc01}. 

\section{The 2000 and 2001 monitoring campaigns}

To date, we have only published the data corresponding to the 
campaigns from 1996 to 1999 (Oscoz et al. 2001). A summary of 
the last two observing campaigns (10/1999 -- 06/2000 and 
09/2000 -- 06/2001) are shown in Table 1. (full 
data corresponding to all the observing campaigns --dates, 
brightness and errors of the individual data-- can be found at 
http://www.iac.es/project/quasar/mserra/meth.html). Each data 
point is the result of averaging several individual measurements. 
The reduction procedure was done by means of the {\it pho2com} 
IRAF\footnote{IRAF is distributed by the National Optical 
Astronomy Observatories, which are operated by the Association 
of Universities for Research in Astronomy, Inc., under 
cooperative agreement with the National Science Foundation.} 
task (for a complete description of {\it pho2com}, see 
Serra-Ricart et al. 1999). Once the final light curves were 
obtained, the data were checked to eliminate inconsistent 
measurements: some points are affected by systematic effects, 
and show strong and {\it simultaneous} (non time-shifted) 
variations in both components. These points are the result of 
bad weather conditions or problems with the CCD and/or the 
telescope. The number of discarded data points was always small 
(18 out of 401).

The final light curves of our monitoring campaign ranging from 
25.02.1996 to 06.06.2001 in the $R$ and $V$ bands are presented 
in Figure 1. The apparent magnitudes of the A and B components 
were derived by comparing the instrumental fluxes with those of 
two reference stars (D and H, see Serra-Ricart et al. 1999). 
From the scatter in the comparison star differential light curve 
we estimate that the photometry is accurate to 2--3 per cent. In 
Figure 1 a delay of 422.6 days has been applied to the B 
component, but no magnitude correction has been applied to the 
data set, both the A and B magnitude are the real ones. It is 
obvious that the behavior of the light curves shows epochs in 
which both components fade, followed by epochs in which they 
brighten, in a quasi-periodic way. This is the general trend 
observed during the 2000 campaign. However, a conspicuous 
behavior can be seen in the 2001 campaign, where a brightening 
of more than 0.2 mag in component A can be observed between day 
$\sim$ JD2451500 and day $\sim$ JD2452065. This behavior is 
evident when only the points corresponding to the 2001 campaign 
are represented. This is shown in the upper panel of Figure 2, 
where the B component is not delayed, and the data have been 
averaged into 10-day bins to reduce the noise and to clearly 
show the trend of both components.

\section{Discussion}

Every year, when the observing season for Q0957+561 is finished, 
the obtained data are reduced by the IAC group together with 
the data from previous campaigns. A possible explanation of the 
trend appearing in the A component during the 2001 campaign is 
that the data have been badly obtained and/or reduced, leading
to a wrong magnitude estimate. However, a mistake in the reduction 
procedure would lead to changes in the whole data, not only in 
the points corresponding to the latest year. Moreover, only the 
A component points show this variation, while the B data remain 
almost constant (see the upper panel of Fig. 2). These facts 
demonstrate that a wrong reduction process or a failure in data 
acquisition can not be the explanation for this trend in the 
image A light curve.

Differential photometry between both comparison stars (see 
Section 2) has been performed in order to check for their 
stability. No significant variability in the differential light 
curve is observed, as can be seen in the lower panel of Figure 2. 
So, the brightening of the A component is certainly not due to 
any change of the reference stars.

Two different explanations for the monotonous increase in the 
brightness of the A component of Q0957+561 are proposed: (i) it 
is intrinsic variability of the source; (ii) it is due to a 
microlensing event, either a short-time one (months to a few 
years) or the cessation of the historical microlensing (about 
twenty years, see Pelt et al. 1998). 

\subsection{Intrinsic variability of the source}

Figure 1 shows that there have been several epochs of remarkable 
intrinsic variability in the last 5 years. However, these 
changes are always less than 0.15 mag; for example $\sim 0.14$ 
mag between JD 2450300 and JD 2450500 or $\sim 0.12$ mag between 
JD 2451100 and JD 2451300. Note that the sharp drop 
detected by \citet{kun95} had an amplitude of 0.13 mag. So, if 
the trend found in the 2001 observing campaign is the consequence 
of intrinsic variability of the source, it would be the largest 
intrinsic variation ever found, with an optical flux increase of 
at least 0.2 mag. The large brightening now detected would make 
it relatively easy to obtain a final confirmation of the time 
delay between both components of Q0957+561. In addition, and 
perhaps even more important, it would allow to obtain this delay 
independently of the method selected, finally solving the 
controversy of the last few years. Thus, a monitoring of Q0957+561 
until 2004 would be crucial to improve our knowledge of both the 
time delay and the robustness of several statistical methods.

\subsection{Microlensing}

As stated before, a microlensing event of more than ten years is 
being produced in Q0957+561. \citet{pel98} made a statistical 
analysis of the Q0957+561 light curves from the first 17 years 
(1979--1996) with data from Schild et al. 
(http://cfa-www.harvard.edu/$\sim$rschild), Princeton University 
(Kundic et al. 1997) and the IAC group first observing campaign 
(Oscoz et al. 1996). These observational data led them to obtain
a time delay of 416.3$\pm$1.7 days, and then to calculate the 
differential light curve between both components of Q0957+561 
(taking into account this value for the time delay). Pelt et al. 
(1998) finally concluded the existence of a first variation of 
0.25 mag in about 6 years, followed by a quiet phase of about 8 
years without variability over 0.05 mag. The historical differential 
light curve is presented in Figure 3, where only data from Schild 
et al. and the first four campaigns of the IAC group campaigns 
have been used. A time delay of 422.6 days has been applied to 
B component data, and only the annual averages are presented. 
The analysis made by Pelt et al. (1998) shows that objects with 
mass of $< 10^{-5} M_\odot$ can explain the 0.25--mag event. They 
also stated that the existence of objects with a mass as high as
$1 M_\odot$ was possible, although they are quite unlikely. A 
remarkable fact since the beginning of this event is that 
component B remains brighter than component A. However, the 
differential light curve does not clearly lead to the 
interpretation of long term microlensing. The shape of this curve 
does not match the one expected for a microlensing event, and 
it is difficult to explain the dip between day 5000 and day 6000.
A point favouring the historical microlensing interpretation is 
that component A is brighter than B in the line emission
(Angonin-Willaime \& Vanderriest 1995), 
which is supposed to be not affected by microlensing.

In any case, a point against the explanation of the observed 
variability in the 2001 campaign as the end of the historical 
microlensing is the fact that this large timescale microlensing 
took six years ($\sim$ 1983-1988) to vary 0.25 mag, while about 
the same variation has been measured now in only a year. 

Another interesting explanation to this large change in brightness 
is that it can be produced by a short timescale (from several 
months to a few years) microlensing event. Until now, no short 
timescale microlensing event has been completely confirmed in 
Q0957+561, although some observing campaigns with several 
participating observatories have been carried out \citep{col02}.
Even the possible microlensing event reported by \citet{sch96} 
is not entirely convincing. This author, with his own data and a 
time delay of 404 days, found amplitude peaks of 0.05 mag and 
90 days long in the microlensing curves. This phenomenon was 
interpreted as short timescale microlensing due to objects 
with 10$^{-5} M_\odot$ mass.  

Refsdal et al. (2000) employed the microlensing light curve by 
Pelt et al. (1998) to restrict the microlens mass. These authors 
concluded that the lens mass could be restricted to values in the 
interval 10$^{-6} M_\odot$-5$M_\odot$.
Another analysis was performed by Schmidt \& Wambsganss (1998) 
with data in the $g$ band by Kundic et al. (1997, two observing 
campaigns: 1994, December to 1995, May and 1995, November to 
1996, July) and a time delay of 417 days. No variation larger 
than 0.05 mag was found in the differential light curve. Two 
conclusions were derived: (i) MACHOs with masses in the interval 
10$^{-5} M_\odot$-10$^{-3} M_\odot$ can be excluded for a quasar
with a radius less than 10$^{-4} h_{60}^{-1/2}$ pc; and (ii) 
there were no evidence of short timescale events. Lately, 
Wambsganss et al. (2000) added to the previous light curves the
data obtained until 1998 in the same band and with the same 
telescope, detecting again no microlensing with amplitude larger 
than 0.05 mag. They could extend the previous limits, excluding 
an halo only made by MACHOS with masses between 10$^{-6} M_\odot$
and 10$^{-2} M_\odot$ for a quasar with radius less than 
10$^{-4} h_{60}^{-1/2}$ pc.

Finally, Gil-Merino et al. (2001) performed an exhaustive 
analysis of the microlensing signal obtained with the IAC 1996, 
1997, and 1998 observing campaigns in the $R$ band. They 
selected a delay of $\Delta t_{Q0957} = 425\pm4$ days 
(Serra-Ricart et al. 1999). Gil-Merino et al. concluded that:
(i) no 3 months duration and 0.05 mag amplitude events are 
found in the microlensing light curves, so these events do not
occur in a continuous way; (ii) from a conservative point of 
view, the amplitude of any microlensing signal must be in the
interval [-0.05 mag, +0.05 mag] (the same limit found by 
Pelt et al. 1998 and Wambsganss et al. 2000); and (iii) the 
small variability observed in the differential light curves 
could be originated, in a natural way, by observational noise 
mechanisms. 

\section{Final remarks}

We present in this letter a large, $|\Delta m| \geq 0.2$-mag, 
brightening of component A of the gravitational lens system 
Q0957+561. The event occurs between day $\sim$ JD2451500 and day 
$\sim$ JD2452065, our last observing date, so its amplitude could be 
even larger. Two different alternatives are offered to explain 
this variation: intrinsic variability or microlensing. 

The historical light curve of Q0957+561 presents several large 
variations in amplitude. Some of them are fast, as the 0.13--mag 
sharp drop detected by \citet{kun95}, whereas others, 
larger in magnitude (but always below 0.15 mag), are relatively 
slow (see Figure 1). However, the detected 2001 variability in 
component A, if intrinsic, would be the largest one ever reported 
in this quasar, allowing so to obtain a confirmation of the time 
delay independently of the method employed. 

Alternatively, the observed brightening could be due to a 
microlensing event. As a possible explanation, it could 
correspond to short timescale microlensing (months to years). 
Microlensing events of this type have been detected in several
gravitational lenses, specially in Q2237+0305, where they are 
almost routinely detected (a noticeable 0.15--mag microlensing 
in component A of Q2237+0305 has been recently reported; 
Wozniak et al. 2000; Alcalde et al. 2002). On the contrary, in 
Q0957+561 it would be the first secure event of this type 
detected (see Schild 1996; Gil--Merino et al. 2001). Another 
possibility is that it could indicate the end of the historical 
microlensing which started in 1983 (Pelt et al. 1996). However, 
the variation seems too fast to correspond to the cessation of 
such a microlensing event.
 
In any case, the definitive answer will only come after the 
observation of component B during 2002. So, an 
exhaustive monitoring of Q0957+561 from several observing groups 
is necessary from now until summer 2002 to study the behavior of 
component B (if the variability is intrinsic, the same behavior 
will appear in B component since $\sim$ JD2452000, April 2001, 
until $\sim$ JD2452600, August 2002), and at least during 2003 
to cover all the event.

\clearpage

\begin{deluxetable}{ccccc}
\tabletypesize{\scriptsize}
\tablecaption{Summary of the 2000 and 2001 observing campaigns of
Q0957+561. 
See http://www.iac.es/project/quasar/mserra/meth.html for more 
details on all the observations.\label{tbl-1}}
\tablewidth{0pt}
\tablehead{
\colhead{Campaign} & \colhead{Filter} & \colhead{Number of points} 
& \colhead{Number of nights} & \colhead{Points removed}}
\startdata
&$R$&138&74&8\\
10/1999-06/2000&&&&\\
&$V$&88&68&3\\
\hline\\
&$R$&92&76&5\\
09/2000-06/2001&&&&\\
&$V$&83&77&2\\
 \enddata
\end{deluxetable}

\clearpage

\begin{figure}
\caption{Upper panel: $R$ band real magnitude light curves of the 
A (filled circles) and B (open squares) data from the Instituto 
de Astrof\'\i sica de Canarias monitoring of Q0957+561. The B data 
have been delayed by 422.6 days, but no magnitude shift has been 
applied between A and B components. The ``real" years (mid date of 
each observing campaign) appear in the top x-axis. 
Second panel: $R$ band data averaged into 10-day bins in order to
reduce the scatter. 
Third panel: same as in the upper panel but in $V$ band.
Lower panel: $V$ band data averaged into 10-day bins.
\label{fig1}}
\end{figure}

\begin{figure}
\caption{Upper panel: The 2000 and 2001 campaigns of Q0957+561 in 
the $R$ band averaged into 10-day bins. A (filled circles) and B 
(not shifted in time, open squares). Note the almost monotonous 
brightening of the A component between days 2500 and 3000. Lower 
panel: Difference light curve of the two comparison stars (H and 
D) to check their stability.\label{fig2}}
\end{figure}

\begin{figure}
\caption{Differential light curve Q0957+561A (t) - Q0957+561B 
(t-$\Delta$ t) of Q0957+561 from Schild's 
and IAC's data. B component data is delayed by 422.6 days. The 
data corresponding to each year have been averaged.
\label{fig3}}
\end{figure}

\clearpage 

\begin{figure}
\plotone{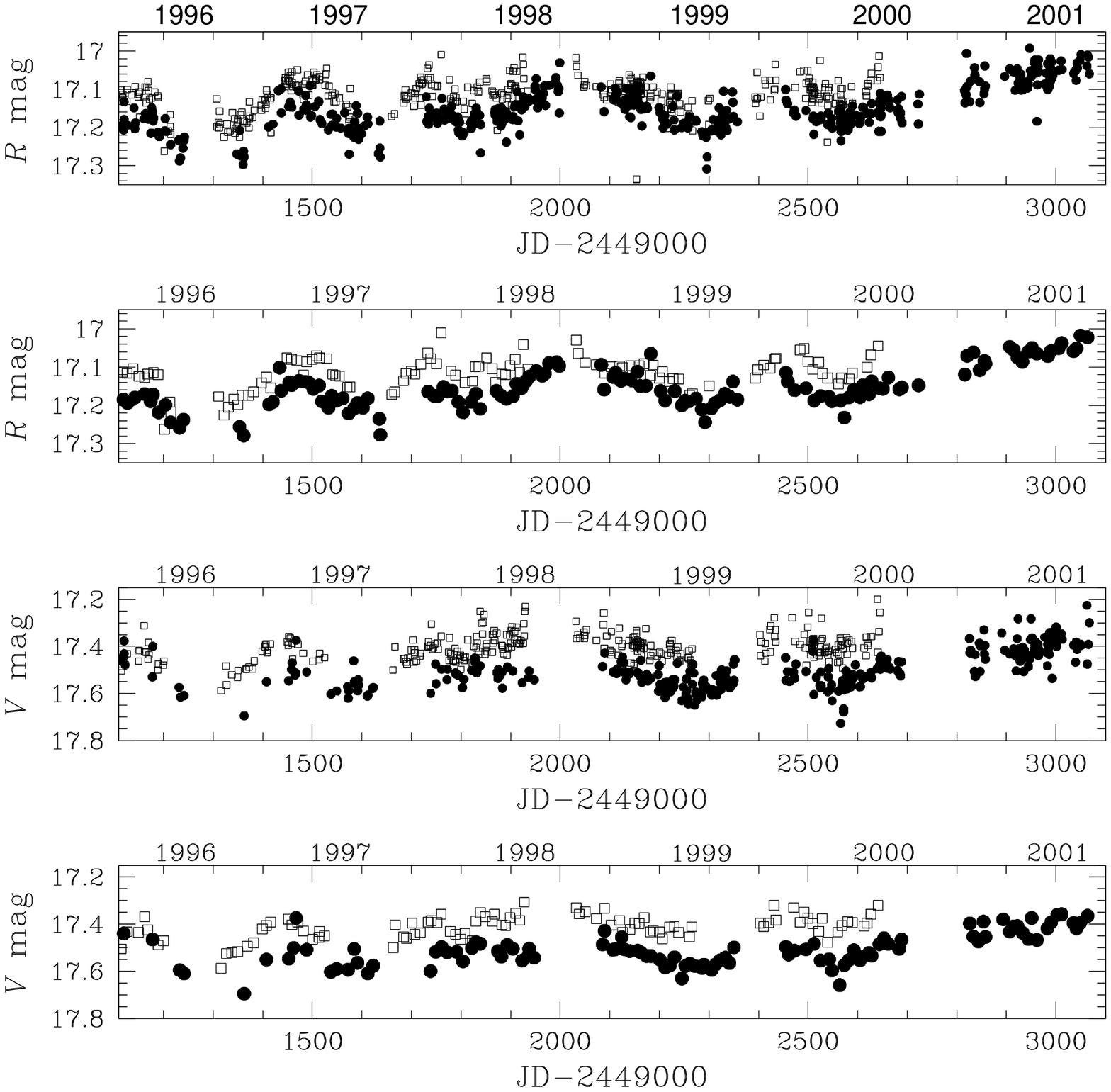}
\end{figure}

\clearpage 

\begin{figure}
\plotone{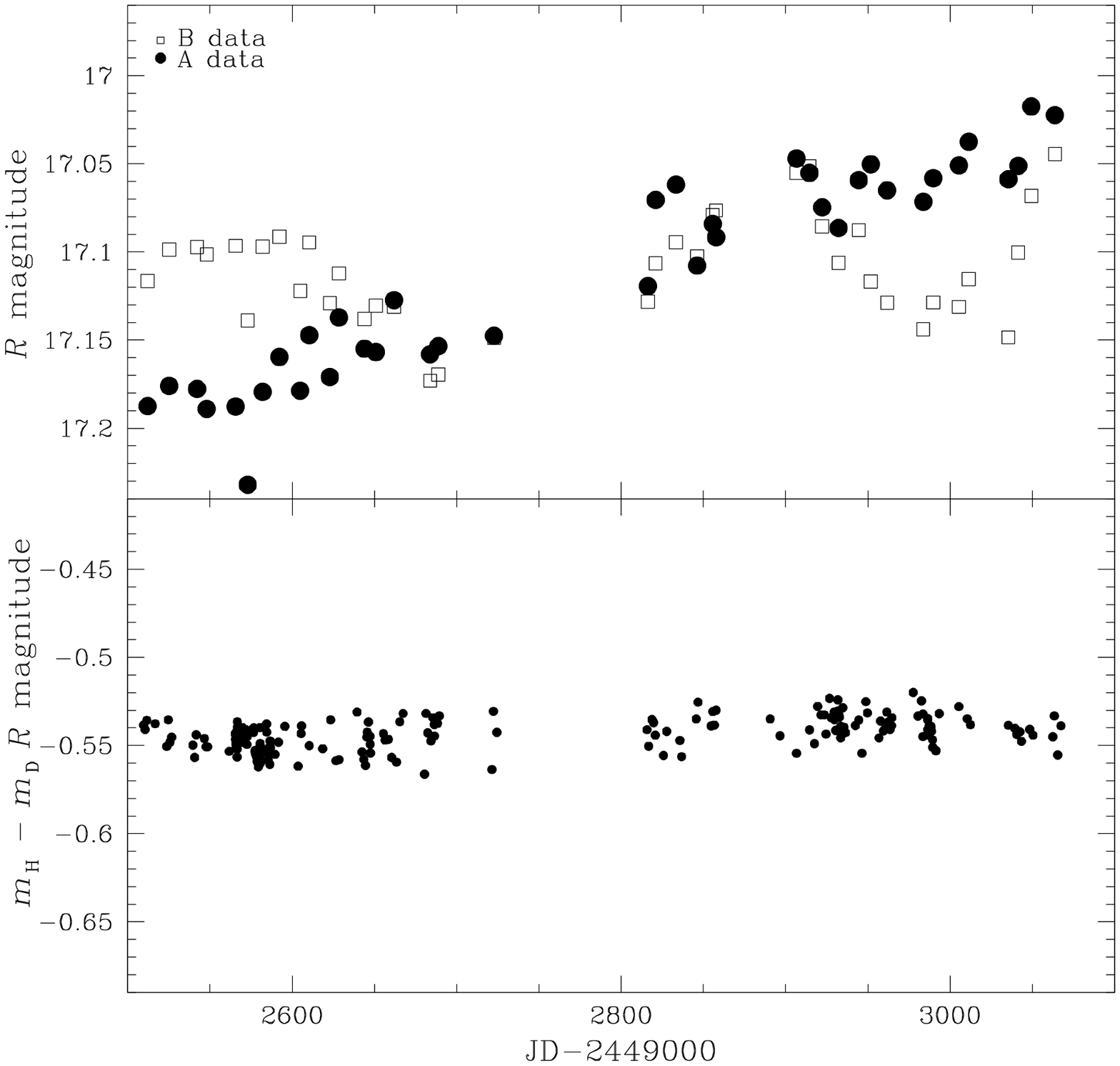}
\end{figure}

\clearpage 
\begin{figure}
\plotone{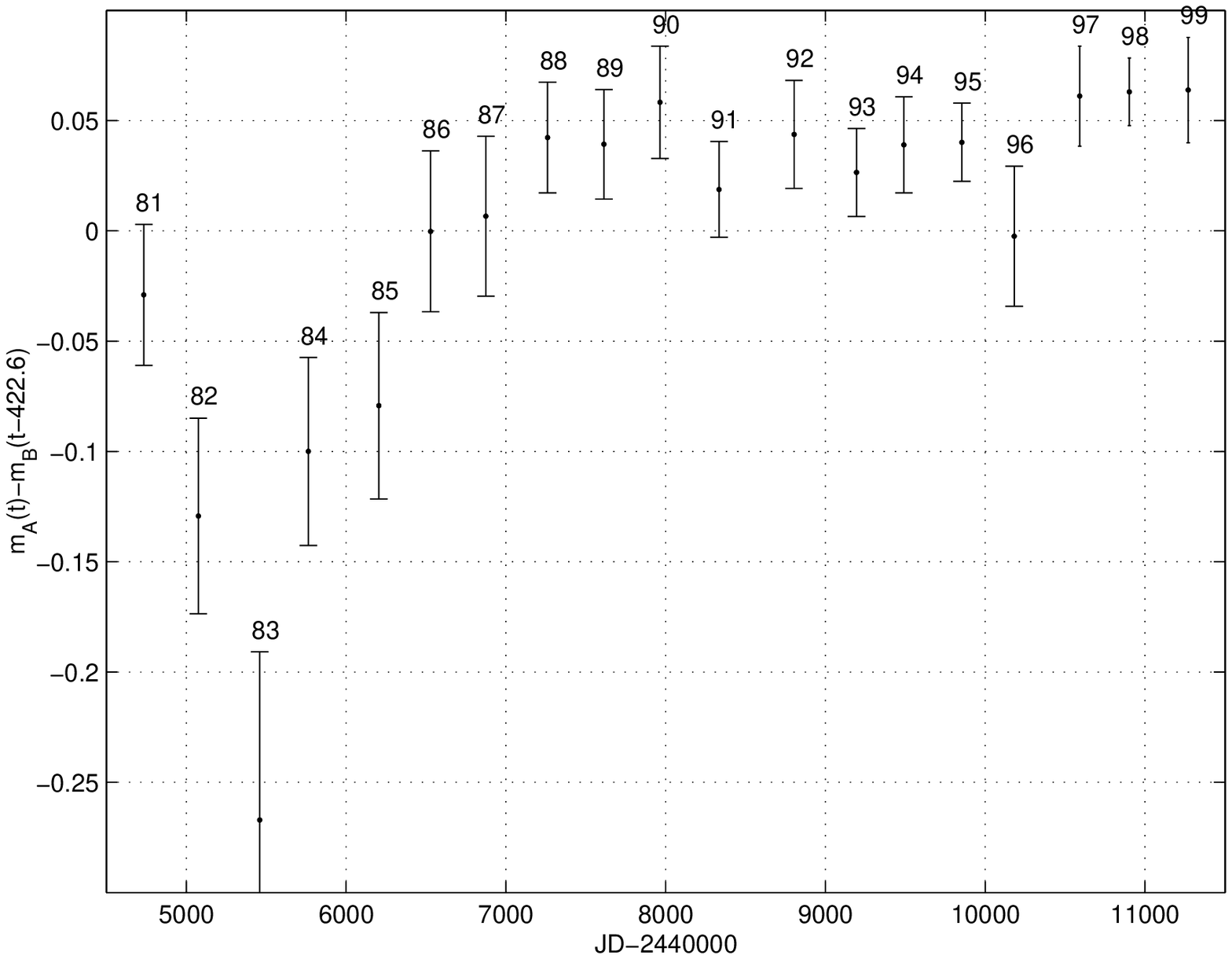}
\end{figure}

\end{document}